\documentclass{phb-proc4-auth}

\usepackage{graphicx}
\usepackage{amssymb}

\begin{document}
\begin{frontmatter}


\journal{SCES '04}


\title{Entropy-Driven Reentrant Behavior in CMR Manganites}

%
%
%
%
%
%

\author[AG]{Nobuo Furukawa\corauthref{1}\thanksref{ACK1}}
\author[RI]{Yukitoshi Motome\thanksref{ACK1}\thanksref{ACK2}}
\author[UT,CE,TS]{Naoto Nagaosa\thanksref{ACK1}\thanksref{ACK2}}

%
 
\address[AG]{Department of Physics, Aoyama Gakuin University, 5-10-1 Fuchinobe,
Sagamihara 299-8558, Japan}
\address[RI]{RIKEN (The Institute of Physical and Chemical Research), 
2-1 Hirosawa, Wako 351-0198, Japan}
\address[UT]{CREST, Department of Applied Physics, University of Tokyo,
7-3-1 Hongo, Bunkyo-ku, Tokyo 113-8656, Japan}
\address[CE]{Correlated Electron Research Center, AIST,
Tsukuba Central 4, 1-1-1 Higashi, Tsukuba, Ibaraki 305-8562, Japan}
\address[TS]{Tokura Spin SuperStructure Project, ERATO,
Japan Science and Technology Corporation, c/o AIST, 
Tsukuba Central 4, 1-1-1 Higashi, Tsukuba, Ibaraki 305-8562, Japan}

%
%
%
%

\thanks[ACK1]{This work is partially supported
 by Grants-in-Aid for Scientific Research
from the Ministry of Education,  
Culture, Sports, Science, and Technology
}
\thanks[ACK2]{This work is partially supported
 by 
NAREGI Nanoscience Project from the Ministry of Education,  
Culture, Sports, Science, and Technology. 
}

%
%
%
%

\corauth[1]{Corresponding Author: Phone: +81-42-759-6292,
Fax: +81-42-759-6549, Email: furukawa@phys.aoyama.ac.jp}


\begin{abstract}

We discuss the origin of the reentrant behaviors of
insulating states above the Curie temperature of the
colossal magnetoresistance manganites.
We consider a system where charge ordering and ferromagnetism
compete with each other. In the presence of randomness
which pins charge order fluctuations, entropy-driven
reentrant behaviors will appear, which explains the
typical temperature dependence of the resistivity
for CMR manganites.

\end{abstract}

%
%



\end{frontmatter}

%
%
%
%
%

\def\TC{T_{\rm C}}
\section{Introduction}

One of the key features of the colossal magnetoresistance (CMR) phenomena
in manganites is that the resistivity shows an insulating behavior 
in the paramagnetic regime above
the Curie temperature $\TC$ and a metallic behavior 
as ferromagnetic moment appears below $\TC$.
By applying an external magnetic field, appearance of the
magnetic moment shifts to higher temperatures 
and the insulating behavior is replaced by the metallic one, 
which creates CMR. 

Experimentally, such behavior has been shown to
be closely related to charge ordering (CO) fluctuations.
Using neutron scattering measurement, Dai {\em et al.} \cite{Dai00} observed
the existence of CO fluctuations above $\TC$ which suddenly disappear
below $\TC$ for the CMR compounds. 
Note that,
for the compounds with less carrier concentrations
 which do now show CMR but are insulating to the lowest temperatures,
 CO fluctuations develop
to make a long-range order (LRO) as temperature decreases.
It has been shown that the temperature dependence of the
CO fluctuations are closely related with that of
the resistivity.

One of the questions which arises from these experimental results is,
provided CO fluctuations are enhanced 
at higher temperatures above $\TC$, why they are disfavored to
make LRO, and instead are taken over by ferromagnetic metal (FM) phase.
In order to clarify the origin of CMR,
it is crucial to understand  this `reentrant' behavior.
Similar phenomena,
where reentrant behavior from CO to FM are observed as decreasing
temperature under a magnetic field, 
have also been reported in (Pr,Ca,Sr)MnO$_3$ \cite{Tomioka2}.

Recently, $A$-site ordered/disordered samples have been synthesized
to make `disorder-control' \cite{Akahoshi}.
In $A$-site disordered samples which exhibit CMR,
Tomioka {\em et al.} \cite{Tomioka1} observed the similar behavior
of CO fluctuations above $\TC$
using X-ray scattering and Raman measurements.
Meanwhile, in $A$-site ordered samples which show the bicritical
phase diagram without CMR, the reentrant behavior is absent.
Thus we see that randomness plays an essential role
to control CMR and reentrant behavior.

The authors have studied the extended double-exchange
system where FM and CO compete
in the presence and absence of quenched randomness, using the  Monte Carlo
technique \cite{Motome03}.
Various phase diagrams as well as 
 disorder-induced insulator-to-metal transition phenomena have been observed.
In this Paper, we further discuss the nature of
the competition between FM and CO 
in various temperature regions,
and investigate the origin of the reentrant behavior of CO fluctuations.

\section{Entropy-Driven Reentrant Behavior}

Let us consider a CO region with insulating behavior
close to the phase boundary to FM region in the absence of disorder.
By adding randomness, CO LRO corrupts 
and becomes short-range order (SRO).
This is due to the anti-phase pinning of the order parameter as
depicted in Fig. 1.
The randomness for conduction electrons corresponds to 
a `random field' to the commensurate CO LRO.

In other words, 
 CO with a solid LRO is energetically unfavorable against
random pinning potentials.
If it is close to the phase boundary, FM becomes relatively
stabilized. Due to the 
competition between these phases, CO disappears
and turns into FM at low temperatures
in the presence of randomness.
We consider that this is the origin of the
disorder-induced insulator-to-metal transition.

At the same time, as shown in Fig. 1, the position of domain boundaries 
can shift without any energy cost, which means that CO-SRO states
have large entropy.
Therefore, at higher temperatures, CO-SRO states will be favored 
by the large entropy due to CO domain configurations, 
which explains the enhancement of CO fluctuations  observed
in our previous report.

This results in an entropy-driven reentrant behavior,
 {\em i.e.}, CO fluctuations become maximum 
above the FM phase.
Such reentrant behavior of the CO fluctuations
is accompanied by the maximum of the resistivity
due to the pseudo-gap structures in the density of states
as well as the optical conductivity which has been reported previously.

\begin{figure}[t]
\centering
\includegraphics[width=7.5cm]{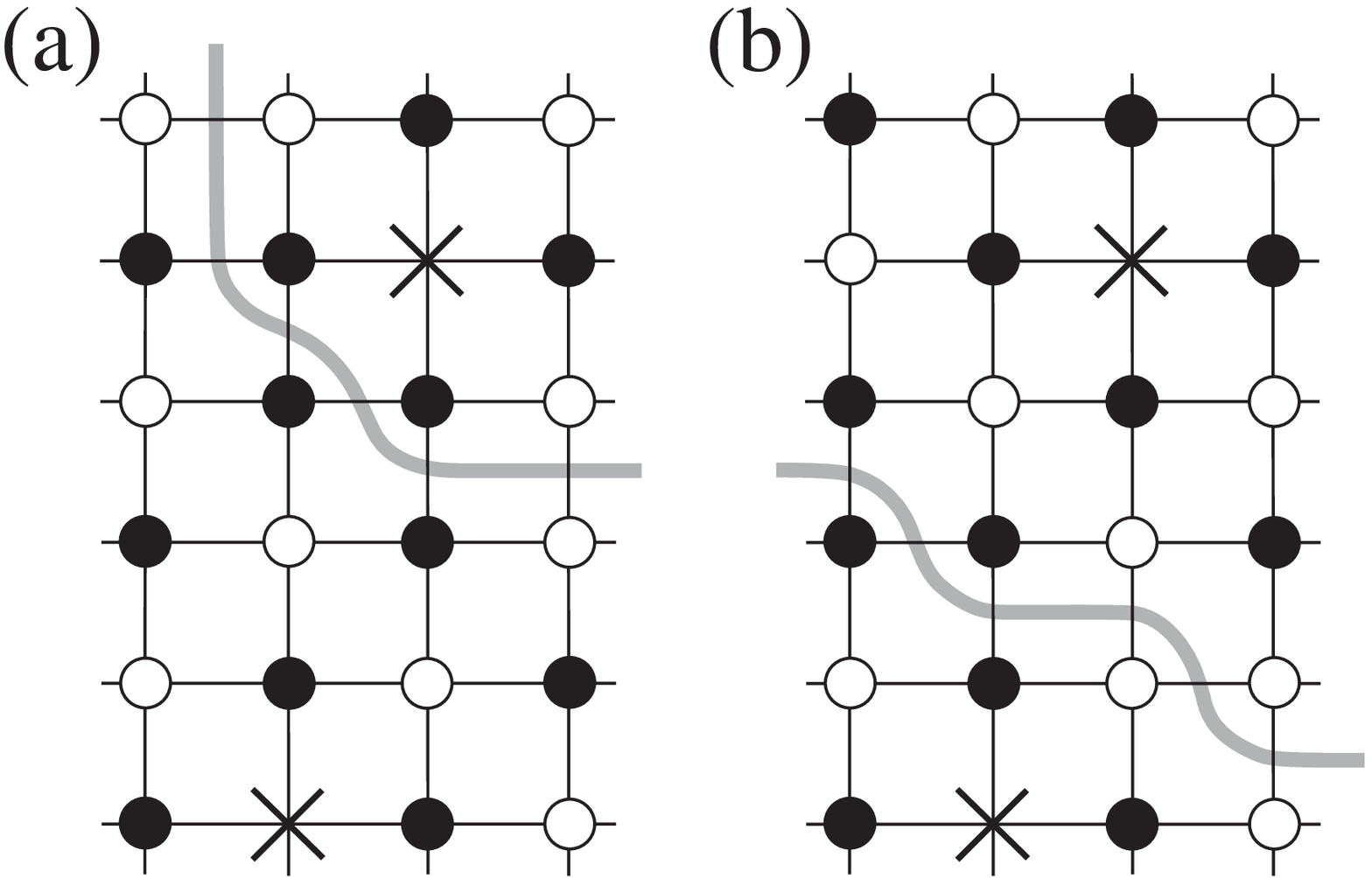}
\caption
{
Schematic pictures of the anti-phase pinning in CO-SRO states. 
White (black) circles denote electrons (holes), and 
crosses are the random pinning centers. 
The dashed gray curves show anti-phase domain boundaries. 
(a) and (b) show different configurations of the domain boundary
for the same configuration of the randomness.
}
\label{fig:ItoM}
\end{figure}

At $T\sim T_{\rm C}$ where maximum competitions between CO and FM
exist, a weak magnetic field stabilizes FM which substantially
reduces the resistivity. We consider that
the competition between CO and FM, as well as
entropy-driven CO-SRO state above FM phase, is
essential to CMR phenomena.

\section{Discussion}

Our proposal for the mechanism of CMR emphasizes the
roles of spacial CO fluctuations with domain structures.
Such  physics will not be captured by
local approximations and mean-field treatments.
In various studies using single-site
dynamical mean-fields, for instance  \cite{Millis,Green},
possibilities for LRO of CO or polaron-lattices have been 
neglected within  theoretical frameworks.
Namely, by not taking into account CO mean-fields,
CO LRO is artificially suppressed within the
self-consistent treatments.
Although the results
of these calculations in the absence of randomness
resemble the experimental results of disordered
samples, they cannot reproduce
the systematic changes between $A$-site ordered and disordered samples.


The authors acknowledge Y. Tokura, Y. Tomioka, and E. Dagotto 
for fruitful discussions. 

%
%
%
%

%
%
%
%


\end{document}